\documentclass[aps,twocolumn, preprintnumbers]{revtex4}
\usepackage{setspace}
\usepackage{graphicx}

\preprint{The following article has been submitted to Journal of Applied Physics}

\begin{document}

\title{Spin lifetime in silicon in the presence of parasitic electronic effects}
\author{Biqin Huang}
\altaffiliation{bqhuang@udel.edu}
\affiliation{Electrical and Computer Engineering Department,
University of Delaware, Newark, Delaware, 19716}
\author{Douwe J. Monsma}
\affiliation{Cambridge NanoTech Inc., Cambridge MA 02139}
\author{Ian Appelbaum}
\affiliation{ Electrical and Computer Engineering Department,
University of Delaware, Newark, Delaware, 19716}

\begin{abstract}
A hybrid ferromagnet/semiconductor device is used to determine a lower bound on the spin lifetime for conduction electrons in silicon. We use spin precession to self-consistently measure the drift velocity vs. drift field of spin-polarized electrons and use this electronic control to change the transit time between electron injection and detection. A measurement of normalized magnetocurrent as a function of drift velocity is used with a simple exponential-decay model to argue that the lifetime obtained ($\approx 2 ns$) is artificially lowered by electronic effects and is likely orders of magnitude higher.
\end{abstract}

\maketitle
\newpage

Manipulation of electron spin injection, transport and detection in semiconductors promises to form the basis of the future's lower-power, higher-speed information-processing paradigm. Much work over the past decade has elucidated useful techniques in this nascent field of spin electronics (or ``spintronics'') but until recently they have been applicable only to direct-bandgap materials. This is either because they are optically-based (electroluminescence,\cite{OHNO, MOLENKAMP, JONKERZNMNSE, HANBICKI} Faraday effect,\cite{KIKKAWARSA, KIKKAWADRAG} or Kerr effect \cite{CROOKER}), or because they require epitaxial (metallic or semiconducting) ferromagnet/semiconductor systems restricted to the few available.\cite{CROWELL, ZEGA} This limitation was particularly unfortunate because it excluded the indirect semiconductor Si, the most dominant semiconductor in the modern microelectronics market, and a particularly attractive material for spintronics due to the expected long spin lifetime.\cite{ZUTICRMP, ZUTICPRL, LYON} For spintronics to deliver on its promise and have a serious hope of integration with conventional information-processing devices,\cite{AWSCHALOMFLATTE} considerable progress must be made in studying and applying spin transport to this material since it forms the basis of the CMOS architecture of computing circuits.

\begin{figure}
   \includegraphics[width=8.5cm,height=8cm]{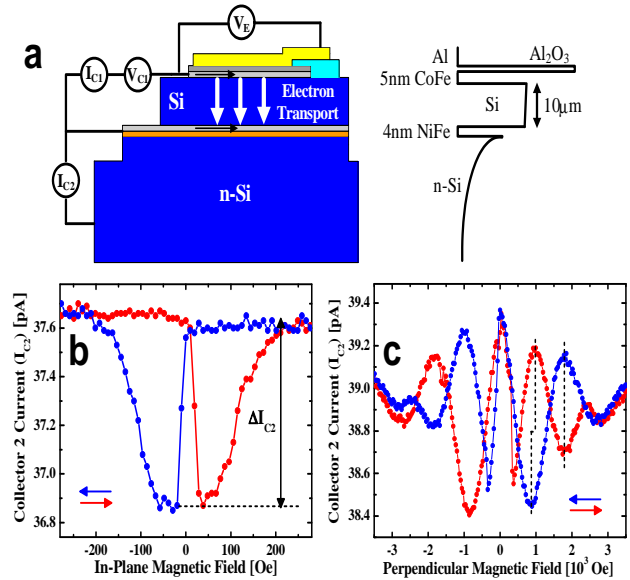}
   \caption{(a)Schematic illustration of the Spin transport device, including side-view and corresponding conduction band diagram. (b) Second collector current, in in-plane magnetic field, showing a spin valve effect at $V_E=-1.8V$, and (c) precession and dephasing (Hanle effect) in a perpendicular field at $V_E=-1.8V$ and $V_{C1}=1.0V$. Note the pronounced shift in the first extrema from zero which is caused by in-plane stray field; see text. }
\label{FIG1}
\end{figure}

Our recent work with spin-dependent hot electron transport in ferromagnetic (FM) thin films for spin injection and detection has solved this long-standing problem and has opened the field of spintronics to include Si, and (in principle) many other indirect-bandgap semiconductors previously excluded from spin transport study.\cite{NATURE} A side-view illustration and a schematic conduction band diagram of our device in equilibrium is shown in Figure 1(a). Hot electrons are emitted from a tunnel-junction (biased with voltage $V_E$) and through a FM thin film to vertically inject spin-polarized electrons into the conduction band of the undoped single-crystal Si layer. This Si is the (100)-oriented device layer of a float-zone-grown silicon-on-insulator (SOI) wafer, so it has very low impurity and defect concentration. After transport through the $10\mu$m-thick Si, the final spin polarization of these electrons is detected on the other side of the Si by ejecting them from the conduction band into a second FM thin film and measuring the magnitude of the ballistic component of this hot-electron current in a n-Si collector on the other side. Typical spin-valve behavior of the second collector current $I_{C2}$ in an in-plane magnetic field is shown in Fig. 1(b), and coherent spin precession and dephasing in a perpendicular magnetic field (Hanle effect)\cite{JOHNSON} is shown in Fig. 1(c). These measurements unambiguously demonstrate coherent spin transport in Si.\cite{MONZON} 
 
In Ref. \cite{NATURE}, we used this type of device to estimate a lower bound on the conduction electron spin lifetime of $\approx$1 ns. Here we establish that electronic effects artificially suppress this lower bound, allowing the actual intrinsic spin lifetime in silicon to be much larger. 

\begin{figure}
   \includegraphics[width=8.5cm,height=8cm]{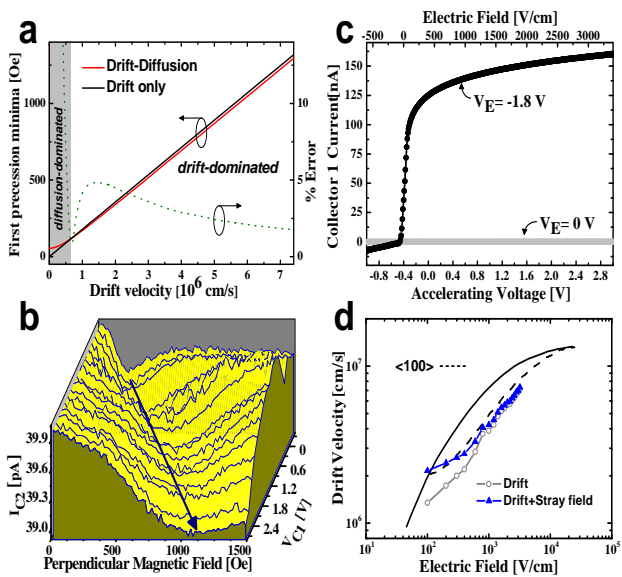}
   \caption{(a) Comparison of predicted prependicular magnetic field value of first minima (corresponding to an average $\pi$ precession angle) using Drift-Diffusion model (Eq. \ref{IC2EQN}) and simple Drift-only model (Eq. \ref{VEQN}). (b) Spin precession measurements in perpendicular magnetic fields at various externally applied accelerating voltages $V_{C1}$. All curves have been shifted to 40pA at zero field, a value typical for injection conditions used ($V_E=-1.8V$ at 85K). The arrow shows the progression of the first minima (used to determine drift velocity) to higher magnetic field values with higher accelerating voltage $V_{C1}$.(c) Injected current $I_{C1}$ as a function of the externally applied accelerating voltage, where the sharp threshold indicates flatband conditions where electric field $\mathcal{E}=0$. (d) Combining the results gained from (b) and (c) allows us to compare with 77K time-of-flight data of charge transport from Ref. \cite{JACOBONI}. Taking into account in-plane stray field (Eq. \ref{QUADRATURE}) gives excellent agreement.}
   \label{FIG2}
\end{figure}

Since we have rectifying Schottky barriers on both sides of the Si drift region, applying a voltage across it does not induce spurious currents. However, it does change the drift electric field and control the drift velocity and hence transit time from injection to detection.  The ratio of second collector current difference ($\Delta I_{C2}$), normalized by the injected current ($I_{C1}$) can therefore be used as a simple voltage-controlled metric which is proportional to spin polarization.\cite{NATURE} (The normalization is necessary because a slight increase in $I_{C1}$ with accelerating voltage increase is caused by enhanced hot-electron collection efficiency at the injection interface).\cite{CROWELLSZE} Clearly, if spintronic effects are to blame for an increase in $\Delta I_{C2}/I_{C1}$ with higher drift velocity, the functional form (not just percent change) should fit to the expected behavior based on exponential spin decay. If it does not, then we must conclude that other effects are blame for artificially suppressing the lower bound derived only from percent change.

As we originally showed in Ref \cite{NATURE}, the effects of precession caused by a perpendicular magnetic field on $I_{C2}$ can be used to determine the transit time $\tau$ from spin injector to spin detector. Because $I_{C2}$ is proportional to the projection of spin direction on the detector thin film magnetization direction, the first minimum in $I_{C2}$ under parallel injector and detector magnetization configuration defines conditions for an average $\omega \tau=\pi$ precession, where $\omega=g\mu_B B/\hbar$ is spin precession angular frequency. To make this interpretation more quantitative, we use the following expression to model the effects of precession in drift (causing drift velocity $v$) and diffusion (causing spin dephasing):\cite{JEDEMA, APPELBAUMSPINFET}

\begin{equation} 
\label{IC2EQN}
\Delta I_{C2} \sim \int_0^{\infty}P(x=L,t)\cos(\omega t)e^{-t/\tau_{sf}}dt,
\end{equation}

\noindent where $P(x=L, t)$ is the solution of the drift-diffusion equation describing the distribution function of electrons arriving at the analyzer (x = L = $10\mu m$ in our device) at time $t$, after spin-polarized injection at $t = 0$.  The exponentially decaying component of the integrand takes into account a finite spin-lifetime, $\tau_{sf}$.

The magnetic field value of the first minima of this expression as a function of drift velocity is shown in Fig. \ref{FIG2} (a). Compared with it is the expected magnetic field dependence in the absence of diffusion, when $\tau =L/v$ so that 

\begin{equation}
v=\frac{2Lg\mu_B B_{\pi}}{h}
\label{VEQN}
\end{equation}

\noindent where $B_{\pi}$ is the perpendicular magnetic field value of the first precession minima in a parallel magnetization configuration. It is clear that two regimes exist: one at higher drift velocity (where drift dominates and the two models predict similar precession minima), and the other at low drift velocity (where diffusion is dominant and the minima value is close to $\approx$ 50 Oe). The relative percent error, given by the dotted line on the right axis, shows that there is little difference between the predictions of the two models in the drift-dominated regime. Since measurements we will show here have minima at perpendicular fields greater than 100 Oe, we conclude that they are firmly in the drift-dominated region. There is therefore negligible error in simply finding the position of the minima of Hanle spectra and using Eq. \ref{VEQN} to determine the drift velocity. 

Typical spin precession measurements at low perpendicular magnetic fields, revealing the first precession minima dependence on accelerating voltage $V_{C1}$, are shown in Figure \ref{FIG2} (b). When the applied voltage increases, the drift velocity increases with electric field and the minima move to higher magnetic field. From the position of these minima, we calculate the drift velocity from Eq. \ref{VEQN} for applied drift-region accelerating voltages between -0.3V and +3.0V.

To correlate these \emph{externally} applied accelerating voltage to \emph{internal} electric drift field, we perform a spectroscopy of injected current $I_{C1}$ as a function of accelerating voltage $V_{C1}$ at constant emitter voltage, as shown in Figure \ref{FIG2} (c). The flat $V_E$=0 V line indicates that in the absence of hot electron injection, no current flows due to the fully rectifying metal-semiconductor interfaces on both sides of the 10$\mu$m Si drift region. However, when hot electrons are injected over the injector Schottky barrier with $V_E$=-1.8 V, a sharp threshold at $V_{C1}\approx -0.4$V is seen. This is where the applied voltage cancels the electric field from ohmic voltage drop in the resistive injector base thin film, and causes flat-band conditions corresponding to zero internal electric field. Since the Si drift region is undoped, the conduction band is linear so that the electric field is constant and linearly related to the voltage difference from the flatband point. Therefore, internal electric drift field is calculated from the externally applied voltage: $\mathcal{E}=(0.4V+V_{C1})/10^{-3}cm$. 

This internal electric field measurement enables us to plot the drift velocity vs. electric drift field. As can be seen with the open symbols in Fig. \ref{FIG2} (d), this data taken at 85K compares well with time-of-flight measurements of charge transport at 77K.\cite{JACOBONI} However, at low fields there is a significant discrepancy. By noting the shift in positions of $\pi$ extrema for parallel and antiparallel magnetization conditions in Fig. \ref{FIG1} (c) (and Fig. 4 (a-c) of Ref. \cite{NATURE}), we conclude that this discrepancy is caused by an in-plane stray field caused by the FM layers. (The shift is clearly not due to measurement delay because it is less apparent for higher-order precession extrema at higher externally-applied perpendicular magnetic fields.) When the injector and detector magnetizations are in a parallel configuration, stray magnetic flux passes through the drift region in-plane, adding in quadrature to the externally applied perpendicular magnetic field, and shifting the minima to artificially low external magnetic field values:

\begin{equation}
B_{internal}=\sqrt{B_{external}^2+B_{stray}^2}.
\label{QUADRATURE}
\end{equation}

When the magnetizations are antiparallel, the stray magnetic field flux is contained by the FM layers and the internal magnetic field equals the external magnetic field. As can be deduced from Fig. 1 (c), this stray field is on the order of $B_{stray}\approx 300 Oe$, and when it is taken into account using Eq. \ref{QUADRATURE}, the correlation between time-of-flight and our corrected data is improved as shown with closed symbols in Fig. \ref{FIG2} (d). 

\begin{figure}
    \includegraphics[width=8.75cm,height=8cm]{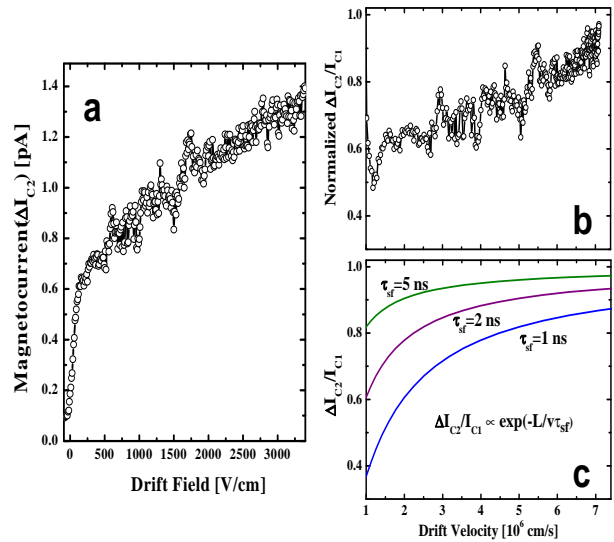}
   \caption{(a) Magnetocurrent $\Delta I_{C2}$ at different electric drift fields from in-plane magnetic field measurements. Note expected threshold at 0 V/cm. (b) Normalized magnetocurrent $\Delta I_{C2}/I_{C1}$ showing linear behavior with drift velocity. (c) Simple exponential-decay model predicting nonlinear behavior of $\Delta I_{C2}/I_{C1}$. The discrepancy between (b) and (c) indicates that the lifetime lower-bound determined from (b) is artificially lowered by electronic effects. }
   \label{FIG3}
\end{figure}

With these electric field and velocity calibrations, we can now attempt to determine $\Delta I_{C2}(v)/I_{C1}(v)$ by measuring $I_{C2}$ in external magnetic fields of -45 Oe and +270 Oe, corresponding to antiparallel and parallel magnetization orientation, respectively (see Fig. 1 (b)). The difference between these measurements is shown in Fig. \ref{FIG3} (a) for a dense set of electric fields up to $\approx$3400 V/cm. This spectroscopy displays the same threshold at flatband conditions that suppresses the injected current $I_{C1}$ (which drives $I_{C2}$) in Fig. \ref{FIG2} (c). The normalized ratio of these two data sets is shown in Fig. \ref{FIG3} (b), where the increase at higher drift field could be assumed to be the result of the correspondingly shorter transit time and hence larger final spin polarization according to $P=P_0exp(-\tau/\tau_{sf})$.

To within the approximation set by our signal to noise level,  $\Delta I_{C2}/I_{C1}$ appears to be a linear function of drift velocity, with the signal magnitude increasing by approximately 50\% from $1\times 10^6$ cm/s to $7\times 10^6$ cm/s. 

To compare with our simple model, we plot $exp(-\tau/\tau_{sf})$, where $\tau =L/v$ for various values of $\tau_{sf}$, on the same drift velocity axis in Figure \ref{FIG3} (c). We ignore the spin injection and detection efficiencies which are unaffected by the electric field within the bulk of the Si, so the vertical scale is normalized for comparison with Fig. \ref{FIG3} (b). To achieve the observed relative 50\% change across the drift velocity range measured in the experimental data, a spin lifetime of approximately 2 ns must be assumed, consistent with our lower bound given in Ref. \cite{NATURE}. However, the nonlinear shape of this modeled curve is in striking contrast to the clearly linear behavior seen in the experimental data.  

What could be the cause? Because the electric field in the drift region changes as the drift velocity (and therefore transit time) change, the spin-orbit interaction which transforms an electric field into a magnetic field could cause deviations from the simple model. However, spin-orbit is very small in Si, and in any case it should act to make the nonlinearity in Fig. \ref{FIG3} (b) stronger, not weaker, in a real measurement. 

Electronic effects such as emitter Schottky field-dependent hot-electron transfer ratio could easily cause the observed linear increase in $\Delta I_{C2}/I_{C1}$, and we believe that these are the most likely source. For instance, as shown in Figs. 2 (b) and 2 (c) of Ref \cite{NATURE}, although $I_{C1}$ drives $I_{C2}$, the relationship is super-linear, making the simple normalization of $I_{C2}$ by $I_{C1}$ insufficient to cancel electronic effects.

In conclusion, we have used coherent spin precession in a perpendicular magnetic field to determine the drift velocity of spin-polarized electrons as a function of applied voltage bias in the drift layer of our Si spin transport device, and used injected current spectroscopy as a function of externally applied voltage to determine the internal electric field. This allows us to then correlate drift velocity and electric field, which agrees well with time-of-flight measurements of electrons in similarly undoped Si.\cite{JACOBONI} By measuring the normalized magnetocurrent $I_{C2}/I_{C1}$ as a function of drift velocity, and comparing to a simple exponential spin-decay model, we determine a lower bound for spin lifetime of free electrons in Si which is limited by electronic effects and is likely orders of magnitude higher.\cite{LYON} Measurements in devices with longer transport lengths will have quadratically higher spin-polarized electron transit times,\cite{APPELBAUMSPINFET} so to probe the 1-microsec regime at fixed accelerating voltage, the drift region should be in the 300-micron range (i.e. the full thickness of a Si wafer).

This work is supported by the Office of Naval Research and DARPA/MTO.

\end{document}